\documentclass[aps,prb,twocolumn,superscriptaddress]{revtex4-2}

\usepackage{graphicx}
\usepackage{bm}
\usepackage{color}
\usepackage{here}
\usepackage{lipsum}

\begin{document}

\title{
Enhanced superconductivity and moderate spin fluctuations suppressed at low energies in heavily electron-doped La1111-based superconductor 
}

\author{T. Kouchi}\email[]{kouchi@nmr.mp.es.osaka-u.ac.jp}
\author{S. Nishioka}
\author{K. Suzuki}
\author{M. Yashima}
\author{H. Mukuda}\email[]{mukuda@mp.es.osaka-u.ac.jp}
\affiliation{Graduate School of Engineering Science, Osaka University, Osaka 560-8531, Japan}
\author{T. Kawashima}
\author{H. Tsuji}
\author{K. Kuroki}
\author{S. Miyasaka}
\author{S. Tajima}
\affiliation{Graduate School of Science, Osaka University, Osaka 560-0043, Japan}

\date{\today}

\begin{abstract}
To elucidate the origin of the re-enhanced high-$T_c$ phase in the heavily electron-doped Fe pnictides, systematic $^{75}$As NMR studies are performed on heavily electron-doped LaFe$Pn$O$_{0.75}$H$_{0.25}$ by controlling the pnictogen height ($h_{Pn}$)  from the Fe plane through the substitution at $Pn$(= As) site with Sb or P.
The measurements of nuclear spin relaxation rate (1/$T_1$) and Knight shift ($K$) reveal that the moderate spin fluctuations at high temperatures are suppressed toward low temperatures. 
Such characteristic spin fluctuations with gap-like feature at low-energies are more enlarged in higher $T_c$ compounds with higher $h_{Pn}$, while those are totally suppressed in non-superconducting compounds with lower $h_{Pn}$. 
This implies that the contribution of the finite-energy part in the spin fluctuation spectrum is crucial for enhancing $T_c$ in the heavily electron-doped regime. This is in contrast to many cases of typical Fe-based compounds with hole and electron Fermi surfaces of similar sizes, where the spin fluctuations at low energies develop significantly at low temperatures.
The features in the heavily electron-doped states are discussed in relation with the characteristics of the faint hole Fermi surface derived from the $d_{xy}$ orbital that rises when $h_{Pn}$ is high, together with the enhanced electron correlation effects. 
\end{abstract}

\pacs{74.70.Xa, 74.25.Ha, 76.60.-k}

\maketitle

\section{Introduction}
Superconductivity (SC) in typical iron pnictides (Fe$Pn$)  LaFeAsO$_{1-y}$F$_{y}$ ($T_c$= 26 K) \cite{Kamihara2008} emerges in the vicinity of a stripe-type antiferromagnetic (AFM) order accompanied by a structural phase transition from tetragonal to orthorhombic (Ort.) phase. The parent electronic states are composed of hole and electron Fermi pockets in similar sizes. 
Remarkable high-$T_c$ states have been discovered in the heavily electron-doped Fe-based compounds, such as single-layer FeSe ($T_c\ge$65 K) \cite{Wang2012,He}, intercalated FeSe systems ($T_c$=30-50 K) \cite{Guo,Y.Mizuguchi2011,A.K.Maziopa2011,A.F.Wang2011,H.D.Wang2011,M.H.Fang2011,T.P.Ying2012,E.W.Scheidt2012,T.Hatakeda2013,L.Zheng2013,T.Noji2014,X.F.Lu2015,S.Hosono2016,M.Z.Shi2018-1,M.Z.Shi2018-2}, and LaFeAsO$_{1-y}$(F/H)$_y$ ($T_c$$\sim$36 K) \cite{Iimura2012,Yang2015}. These high-$T_c$ states appear in the characteristic electronic states composed of large electron Fermi surface (FS) and  no hole FS or the faint hole FS. 
Because neither the magnetic nor the orbital order phases has been reported universally in the vicinity of their SC phases, the indispensable factor for enhancing $T_c$ is still unclear.
Toward  coherent understanding of many Fe-based SCs, it is important to reveal universalities and/or diversities of the SC mechanisms over a wide doping region.

\begin{figure}[htbp]
\centering
\includegraphics[width=10cm]{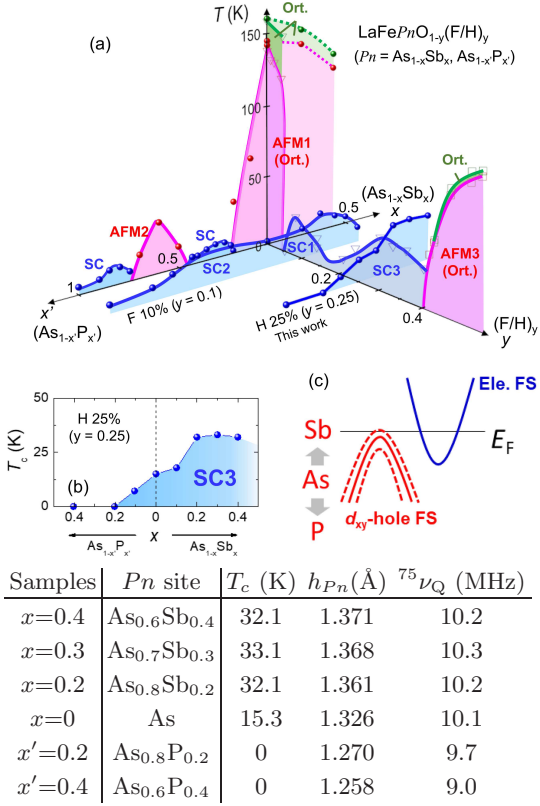}
\vspace{1mm}
\begin{tabular}{c|c|ccc}
Samples & \textbf{$Pn$} site & \textbf{$T_c$} (K) & \textbf{$h_{Pn}$}(\AA) & \textbf{$^{75}$$\nu_{\rm Q}$} (MHz) \\
\hline
$x$=0.4 & As$_{0.6}$Sb$_{0.4}$ & 32.1 & $1.371$ & 10.2 \\
$x$=0.3  & As$_{0.7}$Sb$_{0.3}$ & 33.1 & 1.368 & 10.3 \\
$x$=0.2 & As$_{0.8}$Sb$_{0.2}$ & 32.1 & 1.361 & 10.2 \\
$x$=0 & As & 15.3 & 1.326 & 10.1 \\
$x'$=0.2 & As$_{0.8}$P$_{0.2}$ & 0 & 1.270 & 9.7 \\
$x'$=0.4 & As$_{0.6}$P$_{0.4}$ & 0 & 1.258 & 9.0 \\
\end{tabular}
\caption[]{(Color online)  
(a) Phase diagram of LaFe$Pn$O$_{1-y}$(F/H)$_{y}$  ($Pn$=As$_{1-x}$Sb$_{x}$ and As$_{1-x'}$P$_{x'}$) for $x=0$ \cite{Iimura2012,M.Hiraishi2014}, 0$\leq$$y$$\leq$0.1 \cite{Kamihara2008,Carlsson2011,Hosono2015,H.Luetkens2009,S.Saijo2010,K.T.Lai2014,C.Wamg2009,S.Miyasaka2011,S.Miyasaka2013,S.Miyasaka2017} and $y$=0.25 \cite{S.Miyasaka2017,T.Kawashima}. 
(b) The contents ($x$ and $x'$) dependence of $T_c$ for $y$=0.25 investigated in this study \cite{T.Kawashima}. (c) Schematics of possible band structure made by hole and electron Fermi surfaces around SC3. The theoretical study suggested that the $d_{xy}$ hole FSs around ($\pi, \pi$) increase (reduces) by Sb substitution (P substitution) \cite{K.Kuroki2009}.
The table summarizes the properties of LaFe$Pn$(O$_{0.75}$H$_{0.25}$) used in this study. The values of $T_c$ and $h_{Pn}$ are cited from the previous reports \cite{Carlsson2011,S.Miyasaka2017,T.Kawashima}.
The $\nu_{\rm Q}$'s at the $^{75}$As site are obtained in this work.
}
\label{phase diagram}
\end{figure}

Here, we focus on the reemergent high-$T_c$ phase of heavily electron-doped 1111-compounds LaFe$Pn$O$_{1-y}$H$_{y}$ ($Pn$ = As$_{1-x}$Sb$_{x}$, As$_{1-x'}$P$_{x'}$) \cite{Iimura2012,S.Miyasaka2017,T.Kawashima,S.Miyasaka_un}.
As shown in  Fig. \ref{phase diagram}(a), in this series the electron doping level can be broadly controlled by the content $y$ from the lightly doped SC phase (SC1) to the re-enhanced higher SC phase (SC3) in a heavily electron-doped regime \cite{Iimura2012,S.Miyasaka2017,T.Kawashima,S.Miyasaka_un}.
The $T_c$ within the SC3 phase is enhanced by substitution of the As site with Sb as shown in Fig. \ref{phase diagram}(b), which increases the pnictogen height ($h_{Pn}$) from the Fe plane \cite{T.Kawashima}, whereas the P substitution at the As site decreases $h_{Pn}$ and significantly reduces $T_c$.
Theoretically, it is expected that the variation in $h_{Pn}$ has a significant influence on the energy level of the faint hole FS mainly from the Fe-$3d_{xy}$ orbital \cite{K.Kuroki2009} as shown in Fig. \ref{phase diagram}(c), which increases (decreases) when $h_{Pn}$ becomes high (low), in addition to the electron correlation effects that are more enhanced (degraded) \cite{T.Miyake2010,M.Hirayama2015}.
Therefore, this is a unique system to elucidate the roles of faint hole FS in the vicinity of Fermi level ($E_{\rm F}$)  in the viewpoint of the relationship between the normal electronic state and the SC state.
In heavily electron-doped SC phases such as LaFeAsO$_{1-y}$(F/H)$_{y}$ \cite{Yang2015,H.Yamashita2010,N.Fujiwara2013,N.Fujiwara2015} and intercalated FeSe \cite{H.Kotegawa2011,W.Yu2011,D.A.Torchetti2011,L.Ma2011,Y.Texier2012,Y.P.Wu2015},  spin fluctuations critically enhanced at low energies have not been clearly observed in previous NMR measurements, which is in contrast to the cases of the lightly electron-doped SC states (SC1 \cite{oka2012} and SC2 \cite{shiota2016,sakano2019}), where the spin fluctuations at low energies are significantly enhanced toward low temperatures. 
NMR study is advantageous to elucidate the evolution of the electronic states by means of a common $^{75}$As nuclear probe within a same family compound from the lightly electron-doped states (SC1 and SC2) to the heavily electron-doped state (SC3).

In this study, we report systematic $^{75}$As NMR studies on LaFe$Pn$(O$_{0.75}$H$_{0.25}$) for 0$\leq x\leq$0.4 ($Pn$=As$_{1-x}$Sb$_{x}$) and 0$\leq x'\leq$0.4 ($Pn$=As$_{1-x'}$P$_{x'}$), revealing that characteristic spin fluctuations suppressed at low energies appear explicitly in the Sb-substituted high-$T_c$ compounds, whereas those are totally suppressed in P-substituted non-SC compounds.     
These results suggest that the contribution of the finite energy part in the spin fluctuation spectrum may be rather  important  for enhancing $T_c$ in heavily electron-doped states (SC3), which differs from that of the typical lightly doped Fe-pnictides (SC1/SC2). 

\section{experimental}

Polycrystalline samples of LaFe$Pn$O$_{0.75}$H$_{0.25}$ with nominal contents at $0\le x \le 0.4$  ($Pn$=As$_{1-x}$Sb$_{x}$) and  $0\le x' \le 0.4$  ($Pn$=As$_{1-x'}$P$_{x'}$) were synthesized using a solid state reaction method as described elsewhere \cite{T.Kawashima}. 
Powder x-ray diffraction measurements indicate that the lattice parameters exhibit a monotonic variation with $x$($x'$) \cite{T.Kawashima}.
The bulk $T_c$ values were determined from an onset of zero-resistivity and diamagnetic response in dc susceptibility measurement \cite{T.Kawashima,S.Miyasaka_un}.
As shown in Fig. \ref{phase diagram}(b), the $T_c$  in the SC3 phase at $y$=0.25 increases to $T_{c}\sim33$ K by Sb substitution ($x\sim 0.4$), whereas the $T_c$ decreases to zero by P substitution at $x'>0.2$ \cite{T.Kawashima}.
The $^{75}$As NMR measurements were performed on coarse-powder samples of LaFe$Pn$O$_{0.75}$H$_{0.25}$ with $x=0,0.2,0.3,0.4$ ($Pn$=As$_{1-x}$Sb$_{x}$) and  $x'=0.2,0.4$ ($Pn$=As$_{1-x'}$P$_{x'}$) for the fixed electron-doping level at $y=0.25$.
The Knight shift ($K$) was evaluated by a narrow central peak of the $^{75}$As NMR spectra for the field direction of the $H_{0}\perp c$ axis. 
The nuclear spin-lattice relaxation rate ($1/T_1$) was measured at $H_{0}$$\sim8$ T, which was determined by fitting a recovery curve for $^{75}$As nuclear magnetization to a multiple exponential function $m(t)=0.1\exp(-t/T_1) + 0.9\exp(-6t/T_1)$. 

\begin{figure}[htbp]
\centering
\includegraphics[width=9cm]{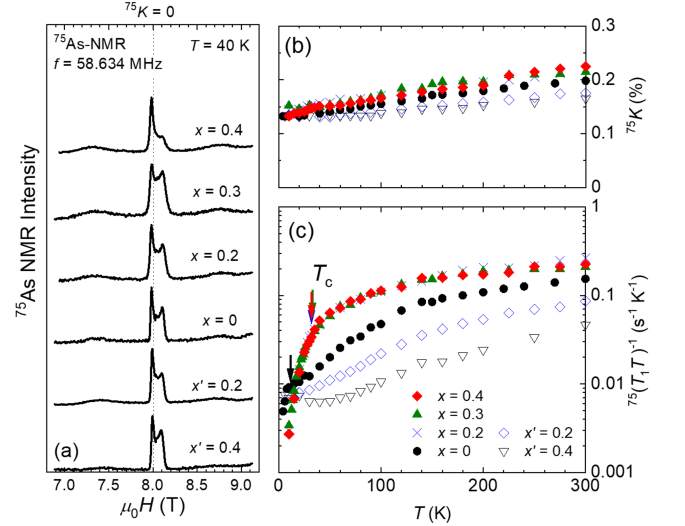}
\caption[]{(Color online)  
(a)  $^{75}$As NMR spectra at $T=40$ K for all the samples, obtained at a fixed frequency 58.634 MHz.
(b) $T$ dependence of $^{75}$$K$ in LaFe$Pn$O$_{0.75}$H$_{0.25}$ for $0\le x \le 0.4$ ($Pn$=As$_{1-x}$Sb$_{x}$) and $0\le x' \le 0.4$ ($Pn$=As$_{1-x'}$P$_{x'}$). 
The $^{75}$$K$ decreases monotonically upon cooling in all the samples. 
(c) $T$ dependences of $^{75}(1/T_1T)$ for each sample are more enlarged in Sb-substituted compounds, although the $T$ dependence of $^{75}$$K$ is almost comparable.
}
\label{spectrum}
\end{figure}

\section{Results and Discussion}

The $^{75}$As NMR spectra of well-defined powder pattern ($I$=3/2) were obtained for all the samples, as shown in Fig. \ref{spectrum}(a).
According to the second-order perturbation theory for the nuclear Hamiltonian with $H_{0}\perp c$ \cite{A.Abragam1961,M.Takigawa1989}, the NMR shifts consist of the Knight shift ($K$) and the second-order quadrupole shift, as expressed by
\begin{equation}
\label{equ1}
\left(\frac{f_0-\gamma_{\rm N}H_{\rm res}}{\gamma_{\rm N}H_{\rm res}}\right) = K + \frac{3\nu^{2}_{\rm Q}}{16(1+K)}\frac{1}{(\gamma_{\rm N}H_{\rm res})^{2}},
\end{equation}
where $\gamma_{\rm N}$ is a nuclear gyromagnetic ratio, $H_{\rm res}$ is a resonance field, and $\nu_{\rm Q}$ is a nuclear quadrupole frequency at the $^{75}$As site.
Here, the electric field gradient asymmetry parameter ($\eta$) is assumed to be close to zero for all the samples in this study, since it was reported previously that $\eta \sim 0$  for  $y<0.3$ in heavily electron-doped LaFeAsO$_{1-y}$F$_{y}$ \cite{Yang2015}.
To evaluate $^{75}$$K$ and $^{75}\nu_{\rm Q}$, the $H_{\rm res}$ was measured as a function of the frequency $f_{0}$.
The slope in the plot of ($f_0-\gamma_{\rm N}H_{\rm res}$)/$\gamma_{\rm N}H_{\rm res}$ against ($\gamma_{\rm N}H_{\rm res}$)$^{-2}$ gives an estimation of $^{75}\nu_{\rm Q}$ for each sample. 
As summarized in the table of Fig. \ref{phase diagram}, the $^{75}\nu_{\rm Q}$ is roughly in proportion to the $h_{Pn}$. Note that the change in $^{75}\nu_{\rm Q}$ is small for $x \ge 0.3$, which may be due to the small change in $h_{Pn}$ \cite{T.Kawashima}.
Since the $^{75}\nu_{\rm Q}$ is proportional to the electric field gradient at the $^{75}$As site, the result ensures the success of monotonic variation of $h_{Pn}$  by the contents $x$($x'$) from the microscopic point of view.

\begin{figure}[htbp]
\centering
\includegraphics[width=8cm]{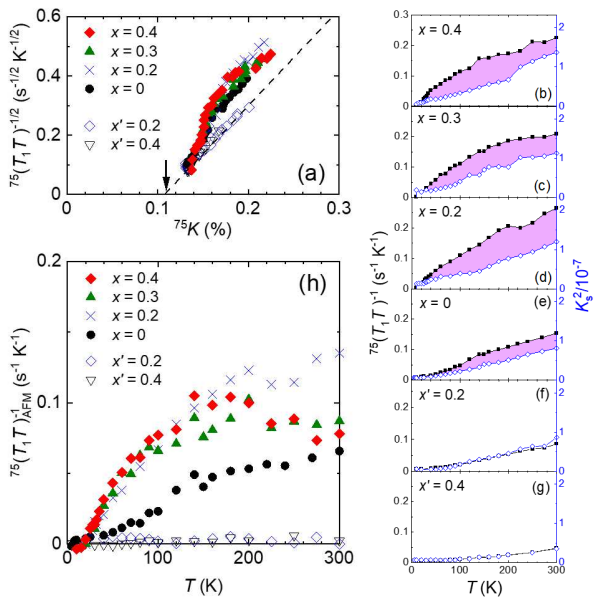}
\caption[]{(Color online) 
(a)  Plot of  $^{75}$(1/$T_{1}T$)$^{0.5}$ vs. $^{75}K$ for LaFe$Pn$O$_{0.75}$H$_{0.25}$ in normal states with an implicit parameter of $T$. 
It is almost linear for $x'$= 0.4, as shown by the broken line, which enables us to estimate $K_{\rm chem}\sim$0.11\%. 
(b)--(g) $T$ dependence of $^{75}(1/T_1T)$ and $K_{\rm s}^{2}$. 
The hatched regions correspond to the component of (1/$T_{1}T$)$_{\rm AFM}$.
(h) $T$ dependence of $^{75}$($1/T_{1}T$)$_{\rm AFM}$, indicating that the AFMSFs appear at high temperature but these are suppressed gradually toward low temperatures, which is more significant in $x\ge 0$ when the $h_{Pn}$ is high. 
}
\label{T1AFM}
\end{figure}

Figures \ref{spectrum}(b) and \ref{spectrum}(c) show the temperature ($T$) dependence of $K$ and $^{75}(1/T_{1}T)$, respectively, for $0\le x \le 0.4$ ($Pn$=As$_{1-x}$Sb$_{x}$) and $0\le x' \le 0.4$ ($Pn$=As$_{1-x'}$P$_{x'}$). The $^{75}K$ decreases upon cooling for all the compounds, which is universal for {\it electron-doped} Fe-based compounds. The $K$ comprises the spin part $K_{\rm s}(T)$ and the $T$-independent chemical part $K_{\rm chem}$, as expressed by $K(T) = K_{\rm s}(T) + K_{\rm chem}$. 
The $K_{\rm s}(T)$ is proportional to uniform susceptibility $\chi(q$=$0)$ with the relation $K_{\rm s}(T) = A_{\rm hf}(0)\chi(q$=$0) \propto A_{\rm hf}(0)N(E_{\rm F})$, where $A_{\rm hf}(0)$ is the hyperfine coupling constant at $q$=$0$, and $N(E_{\rm F})$ is the density of states (DOS) at the Fermi level ($E_{\rm F}$). 
In non correlated normal metals, we expect the relation $(1/T_{1}T) \propto N(E_{\rm F})^{2}$, which corresponds to the Korringa relation expressed as $(1/T_1T)^{0.5}\propto K_{\rm s}$. 
To extract  $K_{\rm chem}$, the $^{75}(1/T_{1}T)^{0.5}$ is plotted against $^{75}K(T)$ for all the compounds, as shown in Fig. \ref{T1AFM}(a).
As for $x'=0.4$, the plot shows almost linear relation (Korringa relation) in the whole $T$ range, indicating that the contribution of the spin fluctuations is negligibly small.
It enables us to estimate $K_{\rm chem}$ to be $\sim$0.11\% for these compounds. 
On the other hand, in the SC compounds of $x \ge 0$, the deviation from the linear relation is clearly seen, indicating that the  $(1/T_{1}T)$ includes additional contribution from the antiferromagnetic spin fluctuations (AFMSFs) at finite wave vectors. 

To extract the component derived from the AFMSFs, we assume that the (1/$T_{1}T$) is decomposed as  (1/$T_{1}T$) = (1/$T_{1}T$)$_{\rm AFM}$ + (1/$T_{1}T$)$_{0}$, according to the previous studies \cite{F.L.Ning2010,shiota2016,H.Mukuda2014-1,H.Mukuda2014-2,M.Miyamato2015}.
The first term (1/$T_{1}T$)$_{\rm AFM}$ represents the component of AFMSFs enhanced at the finite wave vectors (${\bm q}$) at low energies ($\omega \to 0$), which is described as
\begin{equation}
\label{eq2}
\left(\frac{1}{T_{1}T}\right)_{\rm AFM} \propto \lim_{\omega \to 0}\sum_{\bm q}A_{\rm hf}(\bm q)^{2}\frac{\chi^{\prime\prime}(\bm q,\omega)}{\omega},
\end{equation}
where $A_{\rm hf}(\bm q)$ is the hyperfine-coupling constant at $\bm q$ and $\chi^{\prime\prime}(\bm q,\omega)$ is dynamical spin susceptibility at finite wave vector $\bm q$ and energy $\omega$. 
The second term  (1/$T_{1}T$)$_{0}$ is the component related to $N(E_{\rm F})^{2}$, or $K_{\rm s}^{2}(T)$ [= ($K-K_{\rm chem})^{2}$].
Figures \ref{T1AFM}(b)--\ref{T1AFM}(g) show the $T$ dependence of $^{75}$(1/$T_{1}T$) and $K_{\rm s}^{2}(T)$ for each sample. 
Since the $K_s(T)$ shows monotonous decreases upon cooling, the hatched regions in this figure correspond to the component of $(1/T_{1}T)_{\rm AFM}$.  
Consequently, the $T$ dependence of $^{75}(1/T_{1}T)_{\rm AFM}$ is summarized in Fig. \ref{T1AFM}(h). 
We note that in the high-$T_c$ compounds ($x\ge0$) the spin fluctuations develop at the high $T$ region moderately, but these are suppressed gradually below $\sim$100 K.
It indicates that even though the spin fluctuations are dominant in these compounds, the gap like feature appears at low energies in the spin fluctuation spectrum toward low temperatures.  

\begin{figure}[htbp]
\centering
\includegraphics[width=8.5cm]{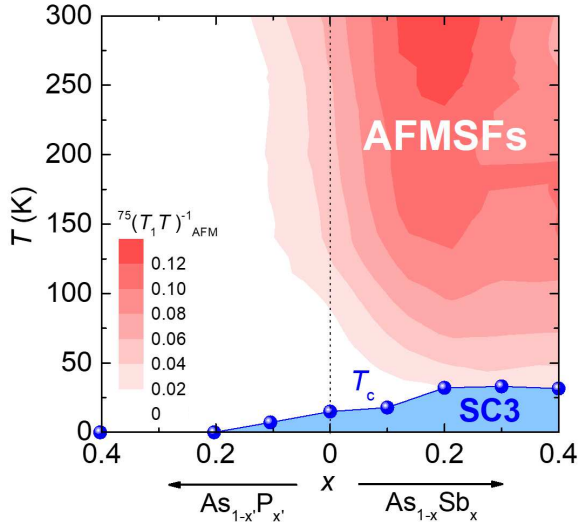}
\caption[]{(Color online)  
Contour plot of $^{75}$($1/T_{1}T$)$_{\rm AFM}$ as functions of $T$ and $x$($x'$) in LaFe$Pn$O$_{1-y}$(F/H)$_{y}$ ($Pn$=As$_{1-x}$Sb$_{x}$ and As$_{1-x'}$P$_{x'}$). The horizontal axis also corresponds to the $h_{Pn}$(see the table in Fig. 1). 
The spin fluctuations exhibit a moderate peak at high temperatures and a suppression toward low temperatures when the $h_{Pn}$ is high, i.e., $x$ is large. 
This behavior of spin fluctuations is more significant in $x\ge 0$ with higher $h_{Pn}$, whereas it is largely suppressed in $x'\ge0.2$ with low $h_{Pn}$, suggesting a possible relation with the reenhancement of $T_c$ in the heavily electron-doped SC3 phase. 
}
\label{T1-SC}
\end{figure}

To reveal the relationship between the $h_{Pn}$, the characteristics of spin fluctuations, and $T_c$, the contour plot of $^{75}$(1/$T$$_1$$T$)$_{\rm AFM}$ is shown in Fig. \ref{T1-SC}, as functions of content $x$($x'$) and $T$. The plot reveals explicitly that the spin fluctuations enhanced at high temperatures are more significant for the higher $T_c$ compounds with higher $h_{Pn}$ (larger $x$), while those are largely suppressed in non-SC compounds with lower $h_{Pn}$ ($x'\ge0.2$).
It suggests that such characteristic spin fluctuations suppressed at low energies are related with the enhancement of $T_c$ in the heavily electron-doped SC3 phase. The suppression of low energy spin fluctuations at low temperatures was also reported in the high-$T_c$ region of heavily electron-doped LaFeAsO$_{1-y}$F$_{y}$ \cite{Yang2015}, which is similar to our results.
This feature differs from that of the SC1 and SC2 phases of \textit{lightly} electron-doped La1111, where the spin fluctuations at low energies are significantly developed toward low temperatures, which has a relation with the enhancement of $T_c$ \cite{oka2012,shiota2016}. 
The significant difference of SC3 in the heavily electron-doped regime can be attributed to the unique FS topology  dominated by the enlarged electron FS and faint hole pocket. 
The band calculation points out that the energy level of the $d_{xy}$ orbital that forms the faint hole pocket also increases  when the $h_{Pn}$ increases \cite{K.Kuroki2009}, which suggests a possible relation with unique spin fluctuations that appears at $x>0$.  
In this FS topology, it is theoretically suggested that the finite energy spin fluctuations give a pairing glue effectively in the heavily electron-doped SC states \cite{M.Nakata2017,K.Matsumoto2020}, even in the absence of well-nested FSs and low energy spin fluctuations. 
Here we also note another important feature, that is, stronger electron correlation effect derived from the higher $h_{Pn}$ \cite{T.Miyake2010,M.Hirayama2015}, which is anticipated  in the SC3 phase as well. 
In the case of heavily electron-doped Ba(Fe$_{1-x}$Co$_{x}$)$_2$As$_2$ \cite{F.L.Ning2010},  both superconductivity and spin fluctuations disappear at $x>0.15$, because $h_{Pn}$ becomes \textit{low} when Fe is replaced with Co \cite{S.Drotziger2010}.
Generally, in the unconventional SCs caused by strong correlation effects, it has been discussed theoretically that the lack of the low energy spin fluctuations that gives rise to a pair breaking may be rather favorable \cite{A.J.Millis1988,Kuroki2005,M.Nakata2017,Matsumoto2018,K.Matsumoto2020,Rodriguez2021}.
Thus, we suggest that one possible reason for the enhanced $T_c$ in the heavily electron-doped SC3 phase is the finite energy spin fluctuations optimized when the FSs are not nested, if the spin fluctuations are a unique factor related to the SC mechanism.

Finally, we compare the feature of the SC3 phase with that of the lightly electron-doped SC1 and SC2 phases.
It should be noted that the P substitution effect on $T_c$ of SC3 is quite different from that in the SC1 and SC2 phases. As shown in Fig. \ref{phase diagram}(a), the SC3 phase disappears drastically by P substitution ($x'>0.2$) while the SC1 phase is robust and the $T_c$ is enhanced again in the SC2 phase at further P substitution \cite{S.Saijo2010,K.T.Lai2014,S.Miyasaka2011,S.Miyasaka2013}.
The SC1 phase is dominated by the hole and electron FSs in similar sizes, which are mainly derived from the $d_{xz}$/$d_{yz}$ orbitals, whereas the composition of the $d_{xy}$ orbital is negligibly tiny since the $h_{Pn}$ is low \cite{K.Kuroki2009}.
The nesting of hole and electron FSs derived mainly from $d_{xz}$/$d_{yz}$ orbitals becomes better especially around the SC2/AFM2 region, which brings about the spin fluctuations at low energies developed significantly toward low temperatures \cite{H.Mukuda2014-1,H.Mukuda2014-2,S.Kitagawa2014,oka2012,shiota2016}.
It plays an indispensable role for enhancing $T_c$ in the SC1 and SC2 phases \cite{H.Mukuda2014-1,S.Kitagawa2014,H.Mukuda2014-2,oka2012,shiota2016}.
Here, in contrast to the SC3, the contribution of the $d_{xy}$ orbital is expected to be negligible since the energy level is far below $E_{\rm F}$ owing to the low $h_{Pn}$. 
Consequently, although spin fluctuations may be generally important for Cooper pairing in Fe-based compounds over the broad doping region, we suggest that the finite energy component of the spin fluctuations may play an important role for enhancing the $T_c$ in the case of the heavily electron-doped Fe-based compounds. On the other hand, in the SC3 region, it is possible that the other low energy local fluctuations such as the orbital fluctuations could play some roles to contribute to the superconductivity, since the structural transition with the AFM3 phase appears in the vicinity of the SC3 phase. However, there is no experimental results to detect the unknown fluctuations except for spin fluctuations due to the lack of large single crystals. Further spectroscopic experiments that reveal the possible roles of orbital degrees of freedom are necessary for a general understanding of the SC mechanism in various Fe-based compounds over wide doping regions.



\section{Summary}
In summary, the high-$T_c$ phase (SC3)  re-enhanced in the heavily electron-doping on LaFe$Pn$O$_{0.75}$H$_{0.25}$ was investigated by the common $^{75}$As NMR probe by controlling the pnictogen height through the substitution at the $Pn$ site. 
We revealed that the moderate spin fluctuations at high temperatures are gradually suppressed toward low temperatures in the high-$T_c$ compounds with high $h_{Pn}$, suggesting the gap like feature appears at low energy regions in the spin fluctuation spectrum toward low temperatures.  
This behavior is more significant when the $h_{Pn}$ is higher in the higher $T_c$ compounds, whereas it is more suppressed in the lower $T_c$ compounds with lower $h_{Pn}$. 
The comparison with the theories on the FS topology suggest that the key element may be the presence of the faint hole FS derived from the $d_{xy}$ orbital and strong correlation effects, suggesting the importance of the finite-energy AFMSFs in the heavily electron-doped states without the nested FSs. 
A similar phenomenon was recently reported in intercalated FeSe compounds \cite{Nishioka2021} that was dominated by the large electron FS and faint hole FS or no hole FS. 
We suggest that this type of AFMSFs may be a characteristic feature, and one of the indispensable factors for enhancing $T_c$ in  heavily electron-doped SC states in iron based compounds. 
However, in these compounds, the contribution of orbital degrees of freedom is still unclear.
Further general understanding of the SC mechanism in Fe-based compounds over a broad doping region should be completed including the possible roles of orbital degrees of freedom, that should be clarified by systematic spectroscopic experiments as well as NMR probes in the future.

\section*{Acknowledgements}

{\footnotesize 
T.K. is supported by a JSPS Fellowship (Grant No. 21J14053) and the Kato Foundation for Promotion of Science (Grant No. KS-3227). 
This work was supported by JSPS KAKENHI (Grants No. 16H04013 and No. 18K18734), the Murata Science Foundation, the Mitsubishi Foundation, and the Tanigawa Fund.
}


\end{document}